\documentclass[final,5p,times,twocolumn]{elsarticle}

\usepackage{amssymb}
 \usepackage{amsthm}

\usepackage{dcolumn}
\usepackage{bm}
\usepackage{color}

\newcommand{\beq}{\begin{eqnarray}}
\newcommand{\eeq}{\end{eqnarray}}

\journal{Physica C}

\begin{document}

\begin{frontmatter}



\title{Superconducting doped topological materials}


\author[label1]{Satoshi Sasaki}
\ead{sasaki@sanken.osaka-u.ac.jp}
\address[label1]{Institute of Scientific and Industrial Research, 
Osaka University, Ibaraki, Osaka 567-0047, Japan}

\author[label2,label3]{Takeshi Mizushima}
\ead{mizushima@mp.es.osaka-u.ac.jp}
\address[label2]{Department of Materials Engineering Science, Osaka University, Toyonaka, Osaka 560-8531, Japan}
\address[label3]{Department of Physics, Okayama University, Okayama 700-8530, Japan}

\begin{abstract}

Recently, the search for Majorana fermions (MFs) has become one of the most important and exciting issues in condensed matter physics since such an exotic quasiparticle is expected to potentially give rise to unprecedented quantum phenomena whose functional properties will be used to develop future quantum technology. Theoretically, the MFs may reside in various types of topological superconductor materials that is characterized by the topologically protected gapless surface state which are essentially an Andreev bound state. Superconducting doped topological insulators and topological crystalline insulators are promising candidates to harbor the MFs. In this review, we discuss recent progress and understanding on the research of MFs based on time-reversal-invariant superconducting topological materials to deepen our understanding and have a better outlook on both the search for and realization of MFs in these systems. We also discuss some advantages of these bulk systems to realize MFs including remarkable superconducting robustness against nonmagnetic impurities.

\end{abstract}


\end{frontmatter}



\section{Introduction}

Majorana fermions (MFs) are neutrinos whose property is that particles are their own 
anti-particles. The existence of MF was initially predicted in high energy 
physics, but has not been experimentally proved yet \cite{Majorana37,Wilczek09}. 
Recently, it has been predicted that the MFs can emerge in topological superconductor 
(TSC) as gapless excitations \cite{Qi11,tanaka12} that are not electrons or holes but 
Bogoliubov quasiparticles formed by bound state due to the superposition of electrons 
and holes \cite{Wilczek09}. The gapless surface or edge state within the bulk 
superconducting gap, regardless of whether or not it is fully gapped or partially 
gapped with nodes \cite{Sato10PRL}, is essentially an Andreev bound state (ABS). It 
is topologically protected since its quantum-mechanical wave function has nontrivial 
topology. After the theoretical prediction, the search for the MFs in condensed 
matter physics has become of great interest \cite{Wilczek09,Schnyder08,qiPRL2009,
Qi10,Sato09,Fu10PRL105,Sato10,Fu08,FuPRL09,Alicea12,Beenakker13,Sau10,satoPRB2009,
SatoPRL09,SatoPRB10,Akhmerov09,tanaka09,Fu10PRL104,mizushima08,mizushimaJPCM2014,
silaev,klinovaja1,klinovaja2} because, if discovered, MFs will deepen our understanding of quantum 
states of matter in physics and foster innovations in future quantum technologies. 
Furthermore, the discovery would provide us a clue to answer what made it 
difficult for us to observe MF as a neutrino.

Since recent progress on research of MFs elucidates that the topological 
superconducting states which harbor MFs can be realized in many different systems 
\cite{Wilczek09,Schnyder08,qiPRL2009,Qi10,Sato09,Fu10PRL105,Sato10,Fu08,FuPRL09,
Alicea12,Beenakker13,Sau10,satoPRB2009,SatoPRL09,SatoPRB10,Akhmerov09,tanaka09,
Fu10PRL104,mizushima08,mizushimaJPCM2014,silaev,klinovaja1,klinovaja2}, it would be useful to look over 
the prerequisites for realizing TSCs. In this review, we focus on bulk 
time-reversal-invariant superconductors (SCs) derived from degenerately-doped 
topological materials that are considered to be a promising platform, namely 
superconducting doped topological insulators (STIs), superconducting doped topological 
crystalline insulators (STCIs), or superconducting heterostructure materials based 
on doped topological insulators (TIs). We will describe advantages/disadvantages of 
those systems that can be distinguished from others in the properties of normal states 
or symmetry invariance. In particular, it is important to notice that theoretically 
the superconducting state in the superconducting doped topological materials is robust 
against non-magnetic impurities \cite{Michaeli,Foster,NagaiPRB,NagaiCon14}. Hence, the 
doped non-magnetic element in topological materials will exert almost no ill effects 
on the realization of a TSC with odd-parity paring state. This is one of advantages 
of superconducting doped topological materials making them to be promising candidates 
for the TSCs.

We discuss the experimental efforts and achievements on the search for MFs 
in superconducting doped topological materials, candidate materials for 
time-reversal-invariant TSCs: Cu$_x$Bi$_2$Se$_3$, Sn$_{1-x}$In$_x$Te, 
(Pb$_{0.5}$Sn$_{0.5}$)$_{1-x}$In$_x$Te, and Cu$_x$(PbSe)$_5$(Bi$_2$Se$_3$)$_6$ 
and list the issues to be addressed in future studies.

\section{Superconducting topological materials}
\label{sec:sti}

TSCs are accompanied by quasiparticles that have nontrivial topological properties defined in the momentum space. The quasiparticles are essentially MFs which are responsible for non-local correlation, non-abelian statistics, and Ising magnetic response~\cite{Wilczek09,Qi11,tanaka12,Alicea12}. 
One of the concrete examples of TSCs is the superfluid $^3$He, which has offered a prototypical system to realize various kinds of topological phases, owing to the richness of symmetry~\cite{mizushimaJPCM2014,silaev,volovik,Sato2014}. As for SCs, only a few bulk materials are known as strong candidates for the host of MFs, e.g., Sr$_2$RuO$_4$~\cite{ueno13} and some heavy fermion SCs, in particular UPt$_3$~\cite{tsutsumiJPSJ2013,saulsAP1994,joyntRMP2002,goswami}.

One of the keys to realize TSCs and MFs in bulk materials is odd parity pairing that satisfies
\beq
P\Delta ({\bm k})P^{\dag} = - \Delta (-{\bm k}),
\label{eq:invD}
\eeq
where $P$ is the inversion operator. This indicates that for a single-band system, only spin-triplet pairing can be a candidate for bulk centrosymmetric TSCs and MFs. A sufficient criterion for realizing TSCs in odd-parity pairing was derived in Refs.~\cite{Sato09,Fu10PRL105,Sato10,sasaki11}, where the topological property is determined by the number of Fermi surfaces that enclose the time-reversal invariant momenta. 

Carrier-doped TIs and topological crystalline insulators (TCIs) which we mainly consider in this paper can offer a promising platform to realize TSCs and MFs for the following two reasons: First, owing to the orbital degrees of freedom of electrons embedded in TIs, odd-parity pairing that satisfies Eq.~(\ref{eq:invD}) can be realized even in an $s$-wave channel of Cooper pairs, as a spin-triplet $s$-wave orbital-singlet, which can be favored by phonon-mediated pairing mechanism. All the odd-parity pairings in doped TIs satisfy a sufficient criterion for realizing TSCs~\cite{Sato09,Fu10PRL105,Sato10,sasaki11}. Second, contrary to conventional wisdom, it has been predicted that a strong spin-orbit coupling makes bulk odd parity superconductivity robust against nonmagnetic disorder~\cite{Michaeli,NagaiPRB,NagaiCon14}. Hence, carrier-doped TIs can expand the horizons of the research field of topological superconducting materials.

In this section, we give an overview of TSCs and time-reversal-invariant TIs as their host materials. Special focus is placed on  Cu$_x$Bi$_2$Se$_3$ as a promising candidate for TSCs. We also make a brief overview on the robustness of bulk odd-parity pairing against disorders~\cite{Michaeli, NagaiPRB, NagaiCon14}. In this paper, we introduce the Pauli matrices in the spin, orbital, and particle-hole spaces, $s_{\mu}$, $\sigma _{\mu}$, and $\tau _{\mu}$, respectively. The repeated Greek indices imply the sum over $x, y, z$ and we set $\hbar \!=\! 1$.

\subsection{$\mathbb{Z}_2$ TIs and Dirac fermions}

To clarify the topological properties of the parent material of STIs, we here start to summarize the discrete symmetries relevant to TIs. Since many time-reversal-invariant TIs hold the inversion symmetry as well as the time-reversal symmetry~\cite{fuPRB2007}, the Bloch Hamiltonian, $\mathcal{H}_{\rm TI}({\bm k})$, satisfies the following relations, 
\begin{eqnarray}
P\mathcal{H}_{\rm TI}({\bm k})P^{\dag} 
= \mathcal{H}_{\rm TI}(-{\bm k}),
\label{eq:inv} \\
\mathcal{T}\mathcal{H}_{\rm TI}({\bm k})\mathcal{T}^{-1} 
= \mathcal{H}_{\rm TI}(-{\bm k}), 
\label{eq:trs}
\end{eqnarray}
where $P \!=\! \sigma _x$ and $\mathcal{T} \!=\! is_y K$ are the inversion and time-reversal operators, respectively ($K$ is the complex conjugation operator). The symmetry~(\ref{eq:trs}) indicates that the eigenstates of the Bloch Hamiltonian, $|u^{(0)}_n({\bm k})\rangle$, have Kramers degenerate partners, $|u^{(0)}_n(-{\bm k})\rangle \!=\! \mathcal{T}|u^{(0)}_n({\bm k})\rangle$.

It is further natural to suppose that $\mathcal{H}_{\rm TI}({\bm k})$ holds a mirror symmetry and $N$-fold rotation symmetry about the $\hat{\bm z}$-axis, because the discrete symmetries originate in crystalline symmetry relevant to TIs: 
\begin{eqnarray}
{M}\mathcal{H}_{\rm TI}({\bm k}){M}^{\dag} = \mathcal{H}_{\rm TI}(-k_x,k_y,k_z),
\label{eq:mirror1} \\
U_N \mathcal{H}_{\rm TI} ({\bm k}) U^{\dag}_N = \mathcal{H}_{\rm TI}(R_N{\bm k}),
\label{eq:rot}
\end{eqnarray}
Without loss of generality, the mirror reflection plane is set to be normal to the $\hat{\bm x}$-axis, where the mirror operator ${M} \!=\! is_x$ changes ${\bm k}$ and ${\bm s}$ to $(-k_x,k_y,k_z)$ and $(s_x,-s_y,-s_z)$. The ${\rm SU}(2)$ matrix $U_N \!=\! \exp(-i\varphi s_z/2)$ describes the $N$-fold spin rotation about the $\hat{\bm z}$-axis by an angle $\varphi \!=\! 2\pi /N$ ($N \!\in\! \mathbb{Z}$) and $R_N$ is the corresponding ${\rm SO}(3)$ rotation matrix. 

Owing to Kramers degenerate band structure, a minimal model for time-reversal-invariant TIs is a $4\times 4$ hermitian matrix. A generic form which holds the symmetries (\ref{eq:inv}) and (\ref{eq:trs}) can be expanded in terms of the five Dirac $\gamma$-matrices as
$\mathcal{H}_{\rm TI}({\bm k}) \!=\! d_0({\bm k}) 
+ \sum _j d_j({\bm k})\gamma _j $~\cite{murakamiScience2003,murakamiPRB2004}.
Following Ref.~\cite{fuPRB2007}, we take the $\gamma$-matrices as
$(\gamma _1, \gamma _2, \gamma _3, \gamma _4, \gamma _5) \!=\! 
(\sigma _x, \sigma _y, \sigma _z s_x, \sigma _z s_y, \sigma _z s_z)$. The symmetries (\ref{eq:inv}) and (\ref{eq:trs}) guarantee that all the coefficients are real and that $d_a({\bm k})$ is even on ${\bm k}$ for $a=0, 1$ and odd otherwise.
The discrete symmetries (\ref{eq:mirror1}) and (\ref{eq:rot}) further constraints $\mathcal{H}_{\rm TI}({\bm k})$ in the lowest order on $k$ as
\beq
\mathcal{H}_{\rm TI}({\bm k}) = c({\bm k}) + m({\bm k})\sigma _x + v_z k_z \sigma _y
+ v ({\bm k}\times{\bm s})_z \sigma _z,
\label{eq:hfinal}
\eeq
where $c ({\bm k}) = c_0 + c_1 k^2_z+c_2(k^2_x+k^2_y)$ and 
$m ({\bm k}) = m_0 + m_1 k^2_z + m_2(k^2_x+k^2_y)$.
The Hamiltonian (\ref{eq:hfinal}) describes the low-energy band structure of topological materials, including Bi$_2$Se$_3$ (TI)~\cite{zhang2009,LiuPRB10}, SnTe (TCI)~\cite{mitchellPR1966,hsiehNP2012,fu}. We also notice that as the lowest order correction to Eq.~(\ref{eq:hfinal}), the effect of the ``warping'' term, $(k^3_++k^3_-)\sigma _zs_z$, was discussed in TIs~\cite{warping} and SCs~\cite{fu14}, where $k_{\pm}\!\equiv\!k_x\pm ik_y$.

The topology of the parent material is characterized by the $\mathbb{Z}_2$ invariant, $\nu$~\cite{fuPRB2007,ando2013}, which originates in the Berry phase of $|u^{(0)}_n({\bm k})\rangle$. The inversion symmetry (\ref{eq:inv}) simplifies the formula for the $\mathbb{Z}_2$ number as
$(-1)^{\nu} \!=\! \prod _i \xi ({\bm \Lambda}_i)$,
where $\xi ({\bm \Lambda}_i) \!= \! \pm 1$ is an eigenstate of $P$ at time-reversal-invariant momenta ${\bm \Lambda}_i$ that satisfy $
[\mathcal{T}, \mathcal{H}_{\rm TI}({\bm \Lambda}_i)] \!=\! 0$. For the form of the Hamiltonian (\ref{eq:hfinal}), the topological invariant $\nu$ is nontrivial (odd) when ${\rm sgn}(m_0m_1) < 0$. Even in the case that $\nu$ is trivial, however, another topological number, the Chern number, can be introduced in a mirror symmetric plane of the Brillouin zone~\cite{fuPRL2011,hsiehNP2012}. A nonzero mirror Chern number defines {\it topological crystalline insulators}. As will be discussed in Sec.~\ref{sec:STCI}, the concrete example is SnTe, which is accompanied by gapless states on the (001) surface, even though the $\mathbb{Z}_2$ number is trivial.

One of the remarkable consequences of an odd $\nu$ is the emergence of topologically protected Dirac fermions that are bound at the surfaces. The wave function of Dirac fermions on the surface ($z\!=\! 0$) is obtained by solving the equation, $[\mathcal{H}_{\rm TI}(k_x,k_y,-i\partial _z)-\mu]{\bm \varphi}_{\rm D}(z) \!=\! \varepsilon(k_x,k_y){\bm \varphi}_{\rm D}(z)$, with a boundary condition ${\bm \varphi}_{\rm D}(0) \!=\! {\bm 0}$. The wave function with the linear dispersion, $\varepsilon (k_x,k_y) \!=\! \pm v\sqrt{k^2_x+k^2_y}$, is given for $c_1\!=\! c_2\!=\! 0$ as
\beq
{\bm \varphi}_{\rm D}(z) = (e^{-\kappa_- z}-e^{-\kappa_+ z})
\left(\begin{array}{c} 0 \\ 1 \end{array}\right)_{\sigma} \otimes u_s(k_x,k_y),
\label{eq:dirac}
\eeq
where 
$\kappa_{\pm}\!=\!v_z/2m_1 \!\pm\!
\sqrt{(m_0+m_2k_{\parallel}^2)/m_1+(v_z/2m_1)^2}$
and $(0,1)^{\rm T}_{\sigma}$ is the spinor in the orbital space ($A^{\rm T}$ denotes the transpose of a matrix $A$). The spinor
$u_s$ in the spin space is determined by $(k_xs_y -
k_y s_x)u_{\pm} \!=\! \pm k_{\parallel} u_{\pm}$.
The penetration depth of the Dirac cone, $\ell$, is defined as $\ell \!\equiv\! \kappa^{-1}_-$. 
The surface Dirac fermions in Eq.~(\ref{eq:dirac}) are fully polarized in the orbital space. 
In Sec.~\ref{sec:pair}, we will discuss the effect of the orbitally polarized Dirac fermions on bulk superconductivity.


\subsection{Topology in superconducting topological materials}
\label{sec:sti}

To derive a sufficient condition for realizing topological superconductivity in doped topological materials, let us here overview the topological invariants, ${\mathbb Z}$ and $\mathbb{Z}_2$ numbers, relevant to time-reversal-invariant SCs. The topological invariant is usually introduced by the global structure of the wave functions of quasiparticles that have a fully gapped excitation spectrum. Such a gapped SC with time-reversal symmetry is categorized to the class DIII in the Altland and Zirnbauer (AZ) table~\cite{Schnyder08,ryuNJP2010}, and the topologically non-trivial property in three-dimension is characterized by a $\mathbb{Z}$ topological number, i.e., a winding number. As clarified in Ref.~\cite{sasaki11}, however, topological invariant can be introduced even in a SC having nodal structures. In this case, the $\mathbb{Z}$ is an ill-defined number due to an ambiguity for redefining the superconducting gap, while a $\mathbb{Z}_2$ that corresponds to the parity of the winding number remains as a well-defined number.

The quasiparticle states that characterize topological superconductivity are obtained by solving the Bogoliubov-de Gennes (BdG) Hamiltonian in the particle-hole space,
\beq
\mathcal{H}({\bm k}) = \left( 
\begin{array}{cc}
\mathcal{H}_{\rm TI}({\bm k})-\mu & -i\Delta ({\bm k})s_y \\
is_y\Delta^{\dag}({\bm k}) & - \mathcal{H}^{\rm T}_{\rm TI}(-{\bm k}) + \mu
\end{array}
\right),
\eeq
where $\mu$ is the chemical potential. In superconducting states, the particle sector is coherently coupled to the hole sector through the pair potential $\Delta ({\bm k})$. Here, we concentrate our attention on time-reversal-invariant SCs that satisfy $\mathcal{T}\Delta ({\bm k})\mathcal{T}^{-1} \!=\! \Delta (-{\bm k})$. The BdG Hamiltonian, then, holds the time-reversal symmetry, $\mathcal{T}\mathcal{H} ({\bm k})\mathcal{T}^{-1} \!=\! \mathcal{H} (-{\bm k})$, and the particle-hole symmetry, $\mathcal{C}\mathcal{H} ({\bm k})\mathcal{C}^{-1} \!=\! -\mathcal{H} (-{\bm k})$ with $C=\tau _xK$. Combining these symmetries, one can define the chiral operator $\Gamma$, which is anti-commutable with $\mathcal{H}({\bm k})$,  
\beq
\left\{ {\Gamma}, {\mathcal{H}}({\bm k}) \right\} = 0, 
\hspace{3mm} \Gamma = -i\mathcal{C} \mathcal{T}.
\label{eq:chiral}
\eeq
The discrete symmetries with $\mathcal{T}^2 \!=\! -1$, $\mathcal{C}^2 \!=\! +1$, and $\Gamma^2\!=\!+1$ 
categorize this system to the DIII
class in the AZ table~\cite{Schnyder08,ryuNJP2010}.

The topological properties of superfluids and SCs are generally determined by the global structure of the Hilbert space spanned by the eigenvectors of the occupied band, $|u_n({\bm k})\rangle$ obtained from $\mathcal{H} ({\bm k})|u_n({\bm k})\rangle \!=\! E_n({\bm k})|u_n({\bm k})\rangle$. To capture the topological property, it is convenient to introduce the so-called $Q$-matrix, 
$Q({\bm k}) \!=\! \sum _{E_n>0} | u_n({\bm k})\rangle \langle u_n({\bm k}) | 
- \sum _{E_n<0} | u_n({\bm k})\rangle \langle u_n({\bm k}) | $~\cite{Schnyder08,ryuNJP2010}.
Since the traceless $Q$-matrix satisfies the conditions, $Q^2({\bm k})=+1$, 
the $Q$-matrix continuously flattens the eigenvalues of $\mathcal{H}({\bm k})$ to $-1$ (occupied) and $+1$ (unoccupied). Therefore, the $Q$-matrix maps the Brillouin zone onto the target space spanned by the eigenvectors of the BdG Hamiltonian. The chiral symmetry is still preserved by the $Q$-matrix, $\{ \Gamma, Q({\bm k})\}\!=\!0$, which is crucial for introducing topological invariant. In the basis that $\Gamma$ is diagonalized to $\Gamma={\rm diag}(+1,-1)$, the $Q$-matrix becomes off-diagonal 
and is reduced to $q({\bm k})\!\in\!{\rm U}(N)$. 
Hence, the relevant homotopy group for the projector $Q({\bm k})$ in three dimensions is given by $\pi _3 [{\rm U}(N)] = \mathbb{Z}$. The topological invariant is defined as the three-dimensional
winding number~\cite{Schnyder08,ryuNJP2010,volovikJETP2009-2},  
\begin{eqnarray}
w_{\rm 3d}=-\int\frac{d{\bm k}}{48\pi^3}\epsilon_{\mu\nu\eta}
{\rm tr}\left[\Gamma (Q\partial_\mu Q)(Q\partial_\nu Q)(Q\partial_\eta Q)\right].
\label{eq:windingQ} 
\end{eqnarray}
For the DIII class, $w_{\rm 3d}$ 
can be an arbitrary integer value, which guarantees the existence of a surface Majorana cone~\cite{Schnyder08,qiPRL2009,volovikJETP2009}. 

As clarified in Refs.~\cite{Fu10PRL105,Sato10,sasaki11}, the $\mathbb{Z}_2$ number that characterizes the parity of $w_{\rm 3d}$ determines a sufficient criterion for realizing time-reversal-invariant TSCs. Here, we suppose an odd-parity pairing that satisfies Eq.~(\ref{eq:invD}). This ensures that the BdG Hamiltonian holds the inversion symmetry up to the $\pi /2$ phase rotation, 
\beq
\mathcal{P}\mathcal{H}({\bm k})\mathcal{P}^{\dag} = \mathcal{H}(-{\bm k}), 
\label{eq:invBdG}
\eeq
where $\mathcal{P}\!=\!P\tau _z$. The inversion symmetry simplifies the parity of $w_{\rm 3d}$ to~\cite{Fu10PRL105,Sato10,sasaki11}
\beq
(-1)^{w_{\rm 3d}} = \prod _{i,m} {\rm sgn}\left[ 
\varepsilon _{2m}({\bm \Lambda}_i)
\right],
\label{eq:z2-1}
\eeq
where $\varepsilon  _{m}({\bm \Lambda}_i)$ is the energy dispersion of the normal state at time-reversal invariant momenta, which is obtained as the eigenvalues of $\mathcal{H}_{\rm TI}({\bm \Lambda}_i)-\mu$ with the chemical potential $\mu$. Owing to the time-reversal symmetry, the energy band of carrier-doped TIs has a Kramers pair as $\varepsilon _{2m}({\bm \Lambda}_i) = \varepsilon _{2m+1}({\bm \Lambda}_i)$. 

The $\mathbb{Z}_2$ number in odd parity SCs was first introduced by Sato~\cite{Sato09} in the case of a single-band system and extended to a general case by Sato~\cite{Sato10} and Fu and Berg~\cite{Fu10PRL105}, independently.
Equation~(\ref{eq:z2-1}) indicates that the mod-2 winding number is determined by counting the number of Fermi surfaces enclosing ${\bm k} \!=\! {\bm \Lambda}_i$. The topological invariant is non-trivial, i.e., $(-1)^{w_{\rm 3d}}\!=\!-1$, when the number of the Fermi surfaces is odd, and trivial otherwise,  $(-1)^{w_{\rm 3d}}\!=\! +1$. Hence, odd-parity pairing and the shape of the Fermi surfaces at time-reversal-invariant momenta are sufficient conditions for realizing topological superconductivity and helical MFs in doped TIs.

Both the $\mathbb{Z}$ and $\mathbb{Z}_2$ numbers require the quasiparticle excitation to be fully gapped, i.e., $\det \mathcal{H}({\bm k}) \!\neq\! 0$. For time-reversal-invariant nodal SCs, however, the node structure is not topologically protected and removed by introducing a small perturbation $\delta$ that transforms $\Delta$ to open a finite gap at the nodal points without breaking the time-reversal symmetry \cite{sasaki11}. The unambiguously-defined $\delta$ is set to be $\delta \!\rightarrow\! 0$ after the $\mathbb{Z}_2$ (not $\mathbb{Z}$) topological invariant is evaluated. This simple procedure enables to define the topological numbers in nodal SCs.

\subsection{Topological odd parity superconductivity in doped TIs (TCIs)}
\label{sec:odd}

Since orbital degrees of freedom is inherent to the parent material of STIs, a Cooper pair potential realized in STIs has orbital degrees of freedom in addition to the spin and momentum dependence. Therefore, a general form of the pair potential in bulk STIs is given by $(i\Delta s_y) _{s_1s_2,\sigma _1\sigma _2}({\bm k})$, where $(s_1,s_2)$ and $(\sigma _1,\sigma _2)$ refere to spin and orbital indices of paired particles, respectively. 
In general, the Fermi-Dirac statistics requires the pair potential to be odd under the exchange of relative coordinates, and orbital states, $(i\Delta s_y) _{s_1s_2,\sigma _1\sigma _2}({\bm k}) \!=\! - (i\Delta s_y) _{s_2s_1,\sigma _2\sigma _1}(-{\bm k})$. For $s$-wave channel, this condition reduces to 
\beq
{\Delta} = s_y {\Delta}^{\rm T} s_y. 
\label{eq:fermi}
\eeq
Here, we suppose even-frequency superconductivity as the bulk pairing of STIs~\cite{tanaka12}. 
In single-band SCs, the condition is satisfied only by spin-singlet even parity pairing states. Owing to orbital degrees of freedom inherent to the parent material of STIs, however, odd parity pairing can be realized even in $s$-wave channel, when the conditions~(\ref{eq:invD}) and (\ref{eq:fermi}) are satisfied.

Let us now specify the possible $s$-wave pairing symmetry that satisfies the conditions~(\ref{eq:invD}) and (\ref{eq:fermi}) for realizing odd parity superconductivity. For spin singlet pairing, the pair potential $\Delta$ that satisfies Eq.~(\ref{eq:fermi}) is generally parameterized with a vector ${\bm d}^{({\rm s})}$ in the orbital space as $\Delta = i\sigma _{\mu} \sigma _y{d}^{({\rm s})}_{\mu} $, which is the spin-singlet even-parity orbital-triplet state. Among the possible orbital-triplet states, only the pairing,
$\Delta \!=\! i\sigma _x \sigma _y{d}^{({\rm s})}_{x} \!\propto\! \sigma _z$, 
can be odd parity pairing. 
A general form of spin triplet $s$-wave pairing is obtained as 
$\Delta \!=\! i s _y s _{\mu} d^{({\rm t})}_{\mu}$,
with a vector in the spin space, ${\bm d}^{({\rm t})}$. This pairing symmetry categorized to the spin-triplet even-parity orbital-singlet state always satisfies the odd-parity condition in Eq.~(\ref{eq:invD}).  

In Table~\ref{table1}, we summarize possible $s$-wave pairing symmetries in STIs with the rhombohedral $D_{3d}$ crystalline symmetry~\cite{Fu10PRL105,sasaki11,yamakage12,hashimoto2013}, which are relevant to both Cu$_x$Bi$_2$Se$_3$ and Sn$_{1-x}$In$_x$Te (See Sec.~\ref{sec:CBS} and \ref{sec:STCI}). There are six independent matrices,  
$(\Delta _{1a},\Delta _{1b}\sigma_x, \Delta _2\sigma_y
s_z, \Delta _3\sigma_z, \Delta _{4x}\sigma_ys_x, \Delta
_{4y}\sigma_y s_y)$, that satisfy the criterion~(\ref{eq:fermi}) for the Fermi-Dirac statistics, where $\Delta _{j}$ ($j$: 1$a$, 1$b$, 2, 3, 4$x$, and 4$y$) denote the amplitude of each pair potential. Their gap structures are also summarized in Table~\ref{table1}, where $(\Delta _{1a},\Delta _{1b}\sigma_x)$ and $\Delta _2\sigma _y s_y$ are fully gapped and the others are accompanied by point nodes~\cite{hashimoto2013,hashimoto14}. We notice that full gap can be realized in the $E_u$ state (see Sec.~\ref{sec:pair}). As mentioned above, all odd parity pairing states can be topologically nontrivial, regardless of nodal structures, when the time-reversal-invariant momentum ${\bm \Lambda} \!=\! {\bm 0}$ is enclosed by the Fermi surface.

A direct consequence of topological odd parity superconductivity is the existence of gapless surface states. The surface structure and tunneling spectroscopy of STIs have been studied theoretically~\cite{yamakage12,hao11,hsieh12,Takami,Yip2013}. The $A_{2u}$ and $E_u$ states in Table~\ref{table1}, which are topological odd parity pairing with point nodes, are accompanied by zero energy flat bands connecting two point nodes that are bound to the surface. We will illustrate in Sec.~\ref{sec:smokinggun} that the zero energy flat band is protected by the hidden ${\bm Z}_2$ symmetry that is obtained by combining the mirror symmetry and time-reversal symmetry. Based on well-established wisdom in unconventional SCs~\cite{TK95,kashiwaya00}, the symmetry-protected surface Fermi arc is responsible for a pronounced zero-bias conductance peak in tunneling spectroscopy. 

\begin{table}
\begin{tabular}{c|c|c|c|c|c}
\hline\hline
$\Delta$ & Gap & Inversion  & Mirror & Topology & $\Gamma$ \\
\hline
$\Delta_{1a}$, $\Delta _{1b}\sigma _x$ & full & $+$ & $+$ & -- & $A_{1g}$ \\
$\Delta_{2} \sigma_y s_z$ & full & $-$ & $-$ & $\mathbb{Z}$ & $A_{1u}$ \\
$\Delta_{3}\sigma_z$  & point & $-$ & $+$ & $\mathbb{Z}_2$ & $A_{2u}$ \\
$\Delta_{4x}\sigma_y s_{x}$ & point &  $-$ & $+$ & $\mathbb{Z}_2$ & $E_u$\\
$\Delta_{4y}\sigma_y s_{y}$ & full &  $-$ & $-$ & $\mathbb{Z}$ & $E_u$\\
\hline\hline
\end{tabular}
\caption{Pairing potentials in STIs, gap structures, and their parity under inversion ${P}$ and mirror reflection ${M} \!=\!is_x$: $\Gamma$ denotes the representation of the $D_{3d}$ symmetry. Note that a full gap behavior in the $E_u$ state is discussed in Sec.~\ref{sec:pair}.
}
\label{table1}
\end{table}

\begin{figure}[t!]
\begin{center}
\includegraphics[width=70mm]{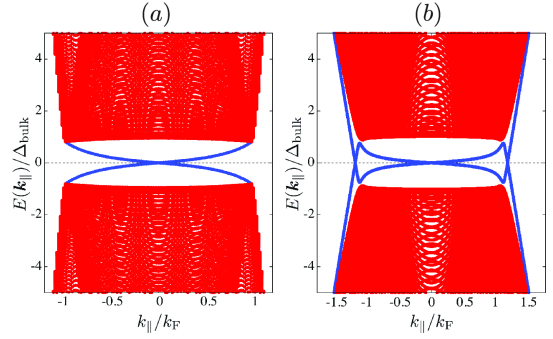}
\end{center}
\caption{Energy spectra of the $A_{1u}$ state: The ``cone'' shape of the surface bound states (blue curve) for $(m_1m_0/v^2_z, m_2m_0/v^2_z) = (-0.59, -0.20)$ (a) and the ``caldera'' shape for $(-0.17, -0.20)$ (b), where $\Delta _{\rm bulk}$ and $k_{\rm F}$ denote the bulk excitation gap and Fermi momentum, respectively.}
\label{fig:fig1}
\end{figure}

The $A_{1u}$ state is similar to the Balian-Werthamer (BW) state realized in the superfluid $^3$He which is a concrete example of time-reversal-invariant TSCs with a full gap~\cite{Schnyder08,qiPRL2009,mizushimaJPCM2014,silaev,volovik}. The BW state is accompanied by the gapless surface state having a linear dispersion, which is called the Majorana cone. The resultant tunneling conductance of the BW state always shows a double-peak structure, rather than a zero-bias conductance peak~\cite{asano03}. However, it has recently been pointed out~\cite{yamakage12,hsieh12} that the dispersion of gapless surface states twists at finite momenta $k_{\parallel} \!\equiv\! \sqrt{k^2_x+k^2_y}$ which is parallel to the surface. The velocity of the Majorana cone near ${\bm k}_{\parallel} \!=\! {\bm 0}$, $\tilde{v}_{\rm surf}$, is given with the parameters in $\mathcal{H}_{\rm TI}({\bm k})$ as~\cite{yamakage12,hsieh12}
\beq
\tilde{v}_{\rm surf} = \frac{1-\sqrt{1+4\tilde{m}_1+4\tilde{m}^2_1\tilde{\mu}^2}}{2\tilde{m}_1\tilde{\mu}^2} \frac{v\Delta}{m_0},
\label{eq:vel}
\eeq
where $\tilde{m}_1\!\equiv\!m_0 m_1 /v^2_z$. At the critical velocity that satisfies $\tilde{v}_{\rm surf} \!=\! 0$, the topological surface state undergoes a structural transition from a ``cone'' (Fig.~\ref{fig:fig1}(a)) to ``caldera'' shape (Fig.~\ref{fig:fig1}(b))~\cite{yamakage12}. 

In Fig.~\ref{fig:fig1}(b), additional crossings of the dispersion of the surface state appear around $k_{\parallel} \!\sim\! k_{\rm F}$. 
The crossing is attributed to the cooperative effect of a remnant of the surface Dirac fermions in the parent TI and the surface MFs~\cite{yamakage12,hsieh12}. Let us suppose the situation that the Dirac cone is well isolated from the conduction band, which corresponds to the situation with low $\mu$ values. This is relevant to Cu$_x$Bi$_2$Se$_3$, where well-defined Dirac cone was observed by angle resolved photoemission spectroscopy (ARPES)~\cite{wray} at the Fermi level. Therefore, the low energy structure is regarded as an effective two-channel model that is composed of the conduction band electrons and Dirac fermions. The conduction band opens a superconducting gap when adding $\Delta _2 \sigma _y s_z$, while the dispersion of the surface Dirac fermions remain gapless near the Fermi energy~\cite{hao11,nagai2013}. Hence, the dispersion (\ref{eq:vel}) of the MFs emergent around ${\bm k}\!=\! {\bm 0}$ must smoothly connect to the dispersion of the gapless Dirac fermions. The cooperative effect of surface Majorana and Dirac fermions necessarily gives rise to the twisting of surface Majorana cone and results in a caldera structure. For a large $\mu$ where the surface Dirac cone merges into the bulk conduction band, the surface states induced by topological odd parity superconductivity remains as a cone shape. The structural transition of the surface bound states occurs in the case of the $E_u$ state as well.

Owing to the structural transition, a pronounced zero-bias conductance peak can be expected not only in the $E_u$ state but also in the $A_{1u}$ state. The $A_{1u}$ state with the caldera shape of surface MFs has a sharp zero energy peak in the surface density of state, which is responsible for the zero-bias conductance peak~\cite{yamakage12,hsieh12,Takami,Mizushima}.

\subsection{Effect of disorders on bulk odd parity superconductivity}
\label{sec:disorder}

As clarified in Sec.~\ref{sec:odd}, odd parity pairing is a sufficient condition for realizing topological superconductivity in doped TIs. Similarly with a conventional $s$-wave pairing, the $A_{1g}$ state in Table~\ref{table1} is highly insensitive to disorder~\cite{Anderson1959}. Contrary to a conventional $s$-wave pairing, however, {\it unconventional} superconducting states have been believed to be fragile against disorders. This suggests that unconventional superconductivity can be realized only in clean materials. Since the superconducting doped topological materials typically have a short mean-free path~\cite{Kriener11}, the robustness of odd-parity superconductivity remains an important issue.

Michaeli and Fu~\cite{Michaeli} recently predicted that in contrast to conventional wisdom, the $A_{1u}$ state, which is one of the odd parity pairings, remains robust against disorder. They found that an approximate chiral symmetry emergent in STIs prevents the superconducting compound from pair decoherence induced by impurity scattering. Similar conclusions were derived in Ref.~\cite{NagaiPRB}, based on a self-consistent $T$-matrix approach for impurity scattering. The robustness against disorder was also reported by Nagai~\cite{NagaiCon14} in the case of the $E_u$ state which has the odd-parity pairing with point nodes.

\section{Candidate materials}

The search for topological phases of matter with time reversal symmetry brought about 
the discovery of TI materials and stimulated the search for an even more exotic state 
of matter, TSC, a superconducting cousin of TI. Opposed to the ideal TIs, the bulk of 
existing materials at the early stage of TI research was conductive and smeared the 
novel properties of the surface states. While achieving a bulk-insulating state in real 
TI samples has been a great experimental challenge, inducing superconductivity in 
TI samples by doping carrier proved to be a great experimental challenge as well.

\subsection{Electron-doped TI: Cu$_x$Bi$_2$Se$_3$}
\label{sec:CBS}

Cu$_x$Bi$_2$Se$_3$, the first example of a superconducting doped TI, is a layered 
material consisting of stacked Se-Bi-Se-Bi-Se quintuple layers that are only weakly 
bonded to each other by van der Waals forces. It has the same rhombohedral crystal 
structure as Bi$_2$Se$_3$ ($R\bar 3m$ space group \cite{Kriener11,Hor10}) with the 
representation of $D_{3d}$ point group \cite{Fu10PRL105}. Its superconductivity 
arises as a consequence of Cu intercalation into the van der Waals gap of the parent 
compound Bi$_2$Se$_3$ yielding an electron concentration of $\sim $2$\times 10^{20}$ 
cm$^{-3}$ with a maximum transition temperature $T_c$ of 3.8 K for $0.12 < x < 0.15$, 
as well as a metallic behavior of the resistivity in the normal state \cite{Hor10}. 
The bulk superconductivity of Cu$_x$Bi$_2$Se$_3$ samples was confirmed by the 
specific heat measurements which showed that the superconducting state of 
Cu$_x$Bi$_2$Se$_3$ for $x \simeq 0.3$ is likely to be fully gapped \cite{Kriener11}. 
The carrier density of order of 10$^{20}$ cm$^{-3}$ is very low for SCs whose maximum 
$T_c$ is in the order of a few K within the framework of the Bardeen-Cooper-Schrieffer 
(BCS) theory. The anomalously high $T_c$ for a relatively low carrier density may 
indicate that the pairing mechanism of the superconducting Cu$_x$Bi$_2$Se$_3$ could 
be affected by the strong spin-orbit coupling, albeit still be governed by the 
electron-phonon mechanism in the same way as most of the existing SCs.

Stimulated by the discovery of superconductivity in doped TI Cu$_x$Bi$_2$Se$_3$, in 
2010, Fu and Berg discussed the possible pairing symmetry within a two-orbital model 
and a sufficient criterion for realizing time-reversal-invariant TSCs in materials 
with inversion symmetry \cite{Fu10PRL105}. Intriguingly, a novel spin-triplet pairing 
with odd parity is favored by strong spin-orbit coupling with gapless surface ABSs 
within a full pairing gap in the bulk: We emphasized in Sec. 2.2 that what is 
important for TSCs is odd parity pairing (See Eq. (\ref{eq:invD})) and a nontrivial 
$\mathbb{Z}_2$number is defined for all odd parity pairing even in nodal SCs.

The discovery of the Cu$_x$Bi$_2$Se$_3$ SC attracted a lot of attention in the 
community, however, experimental studies of this material were hindered at the 
beginning by difficulties of material synthesis. Thanks to employing electrochemical 
Cu intercalation (ECI) technique combined with the post annealing that is essential 
for the activation of the superconductivity (a typical sample image of ECI 
Cu$_x$Bi$_2$Se$_3$ is displayed in Fig. 2(a)) \cite{Kriener11,Kriener11PRB}, the 
maximum shielding fraction of Cu$_x$Bi$_2$Se$_3$ now can reach up to $\sim $50$\%$ 
at 1.8 K with Cu concentration $x$ of $\sim $0.3; besides, the maximum value can be 
even larger at lower temperatures. Meanwhile, serious improvements of the melt-grown 
(MG) synthesis have been done \cite{Lawson12, Maeda, kirzhner12}. Currently 
Cu$_x$Bi$_2$Se$_3$ has been the most widely studied as a candidate for the TSCs.

It would be prudent to discuss disorder effects in Cu$_x$Bi$_2$Se$_3$ since odd-parity 
pairing state in general disfavors disorders \cite{Balian, Larkin, Maeno} as mentioned 
in Sec.\ref{sec:disorder}. There is a systematic unbalance between the number of the 
intercalated Cu atoms and that of the increase in the carrier density \cite{Kriener11}. 
Progressive Cu intercalation into the system introduces significant disorder 
\cite{Hor10,Kriener11PRB,Maeda} and an intrinsic inhomogeneity \cite{Kriener11PRB}, 
leading to an anomalous suppression of the superfluid density which was deduced from the 
measurements of the lower critical field \cite{Kriener12} and an optical spectroscopic 
study \cite{Sandilands}. At the same time, the transition temperature $T_c$ is 
monotonously and only moderately suppressed from $T_c \simeq 3.8$ K ($x \simeq 0.1$) to 
$T_c \simeq 2.6$ K ($x \simeq 0.5$), which agrees with the non-magnetic impurity effect 
theoretically predicted for this class of TSCs \cite{Michaeli, Foster, NagaiPRB, NagaiCon14}. 
The robust topological superconductivity even in disorder can be derived from the 
relativistic effects in doped topological materials \cite{NagaiPRB, NagaiCon14}.

\begin{figure}[t!]
\begin{center}
\includegraphics[width=65mm]{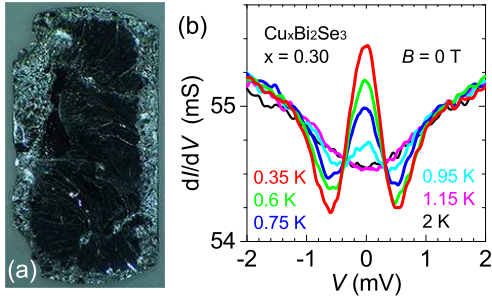}
\end{center}
\caption{Cu$_x$Bi$_2$Se$_3$ and zereo-bias conductance peak. (a) A typical cleaved 
surface image of Cu-intercalated Bi$_2$Se$_3$ with post annealing. (b) Point-contact 
spectra ($dI/dV$ vs bias voltage) of Cu$_x$Bi$_2$Se$_3$ with $x = 0.3$ for 0.35--2 K 
measured in 0 T.}
\label{fig:fig2}
\end{figure}

In order to confirm whether or not a SC is a TSC, it can be useful to verify 
the existence of an ABS consisting of MFs that is a low-energy gapless excitation 
at the surface by detecting surface ``low-energy'' phenomena, in particular electron 
tunneling. The gapless excitations that create the density of states within the 
superconducting gap can be detected as a zero-bias conductance peak (ZBCP) in 
tunneling spectroscopy. Indeed, both the $A_{1u}$ and $E_{u}$ states in 
Table~\ref{table1}, which have bulk topological odd parity superconductivity, are 
responsible for a pronounced ZBCP on the (111) surface of Cu$_x$Bi$_2$Se$_3$ 
\cite{yamakage12,hsieh12}. It should be mentioned that a ZBCP was observed in 
Sr$_2$RuO$_4$, a strong candidate for time-reversal breaking chiral $p$-wave TSCs 
\cite{Maeno,Kashiwaya}.

Actually, the ZBCP was observed by soft point-contact spectroscopy experiments on a 
cleaved (111) surface of ECI Cu$_x$Bi$_2$Se$_3$ for $x \simeq 0.3$ with a tiny drop 
of silver paint consisting of clusters of Ag grains that produce Ag-grain parallel 
channels at the point contact as shown in Fig. 2(b) \cite{sasaki11}. A typical 
diameter of grains appears to be $\sim $20--50 nm estimated from the image taken by 
the atomic force microscope in Fig. 3(a) and the diameter of the contact area for 
one grain $d$ is always smaller than that of the grain size (see Fig. 3(b)). In the 
case of ballistic electron transport, the resistance of the channel can be estimate 
as Sharvin resistance that is proportional to $1/d^2$ while in the case that 
electrons are scattered and loose kinetic energy, the channel resistance can be 
estimated as Maxwell resistance that is proportional to $1/d$ \cite{Gonneli}. The 
Sharvin resistance becomes dominant when $d$ is smaller than the intersection point 
$d_\textsc{I}$ between two types of resistance with respect to $d$ as indicated in 
Fig. 3(c), which is $\sim $100 nm in the case of Ag \cite{Tinkham}. Since the Ag 
grain size is smaller than $d_\textsc{I}$, the electron transport in the case of the 
soft point contact can be ballistic, and hence spectroscopy in the presence of a 
finite repulsive potential barrier at the boundary between the Ag grain and a SC 
can be performed. Theoretical considerations of all possible superconducting states 
shows that the observed surface ABS of Cu$_x$Bi$_2$Se$_3$ is due to MFs \cite{sasaki11}. 
Consistently, similar ZBCPs were observed by other point-contact spectroscopy 
experiments with Au tip on MG Cu$_x$Bi$_2$Se$_3$ for $x \simeq 0.25$ \cite{Chen12}.

\begin{figure}[t!]
\begin{center}
\includegraphics[width=65mm]{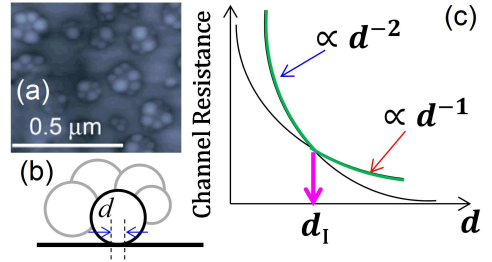}
\end{center}
\caption{Legitimacy of soft point contact. (a) Atomic force microscope picture of 
Ag nano grains on the measured surface after the measurement lead is removed. 
(b) A diameter of the contact area of a nano grain in a Ag cluster at the sample 
surface. (c) An expected diameter dependence in the resistnace of an electron-transport 
channel.}
\label{fig:fig3}
\end{figure}

On the contrary, both unconventional ZBCP and conventional Blonder-Tinkham-Klapwijk 
(BTK) spectra were observed on the surface of MG Cu$_x$Bi$_2$Se$_3$ for $x \simeq 0.20$ 
by point-contact spectroscopy with Au tip \cite{kirzhner12}. Andreev reflection 
spectroscopy performed by a nanoscale device showed spectra partially fitted by BTK 
theory \cite{peng13}. Moreover, a conventional tunneling spectrum was observed on a 
cleaved surface of ECI Cu$_x$Bi$_2$Se$_3$ for $x \simeq 0.25$ by scanning tunneling 
spectroscopy (STS) \cite{Levy13}. These aroused a controversy with respect to the 
nature of superconductivity in Cu$_x$Bi$_2$Se$_3$. It is worth noting that recent 
self-consistent calculations for surface density of states (SDOSs) has clarified this 
puzzling issue: the surface Dirac fermions in a conventional pairing state of the 
$s$-wave SC will open an additional gap larger than the bulk superconducting gap 
yielding a two-gap structure due to parity mixing of the pair potential near the 
surface, nonetheless the STS spectra exhibited only single-gap structure. Importantly, 
the parity mixing is absent in an unconventional pairing state of odd-parity SC 
\cite{Mizushima}. In addition, recent ARPES experiments revealed that the Fermi 
surface (FS) of superconducting Cu$_x$Bi$_2$Se$_3$ is a cylindrical open FS 
\cite{Lahoud} coexisting with the topological gapless surface states of Bi$_2$Se$_3$ 
well-separated from the bulk conduction band \cite{wray,Lahoud}. Therefore the STS 
spectra can be originated from topological odd-parity superconductivity with a 
cylindrical FS \cite{Mizushima}. Further details will be discussed in Sec.~\ref{sec:pair}.

It is worth mentioning that the pressure variation of the superconducting phase of 
MG Cu$_x$Bi$_2$Se$_3$ for $x \simeq 0.3$ was investigated by applying a hydrostatic 
pressure up to 2.31 GPa \cite{Bay12} and the superconductivity appears to be robust 
up to 6.3 GPa. 
The sufficiently large mean free path and the variation of $B_{c2}(T)$ that agrees 
with the $p$-wave polar-state model indicate the $p$-wave spin-triplet 
superconductivity of Cu$_x$Bi$_2$Se$_3$, although an anisotropic spin-singlet state 
cannot be discarded completely. Apparently further studies on Cu$_x$Bi$_2$Se$_3$ 
with different measurement techniques are required to elucidate the true nature of 
this material.

\subsection{Hole-doped TCI: Sn$_{1-x}$In$_x$Te and (Pb$_{0.5}$Sn$_{0.5}$)$_{1-x}$In$_x$Te}
\label{sec:STCI}

Based on a guiding principle suggested from the discovery of Cu$_x$Bi$_2$Se$_3$, 
low-carrier-density semiconductors whose Fermi surfaces are centered around 
time-reversal-invariant momenta in strong spin-orbit coupling systems, in particular 
doped TCIs with mirror symmetry, are other likely candidates for TSCs. In-doped SnTe 
(denoted Sn$_{1-x}$In$_x$Te), a first example of STCIs, has the same rocksalt crystal 
structure ($Fm\bar 3m$ space group) with the representation of $O_h$ point-group 
symmetry as pristine SnTe (see Fig. 4(b)) at ambient temperature. It reveals a metallic 
behavior of the resistivity in the normal state in single crystals synthesized in two 
different ways: Melt growth \cite{Erikson} and vapor-solid growth \cite{SasakiPRL12,
SasakiNanoplate}. A typical faceted single crystal is shown in Fig. 4(a). It has been 
known that hole-doped SnTe, a degenerately-doped TCI due to Sn vacancies, superconducts 
and $T_c$ increases as the hole density increases from 0.024 K ($\sim $5$\times 10^{20}$ 
cm$^{-3}$) to $\sim $0.2 K ($20\times 10^{20}$ cm$^{-3}$) \cite{Hulm68}. When cooled, 
ferroelectric structural phase transition (SPT) occurs in SnTe so that superconducting 
degenerately-doped SnTe must have a rhombohedral crystal structure.

\begin{figure}[t!]
\begin{center}
\includegraphics[width=70mm]{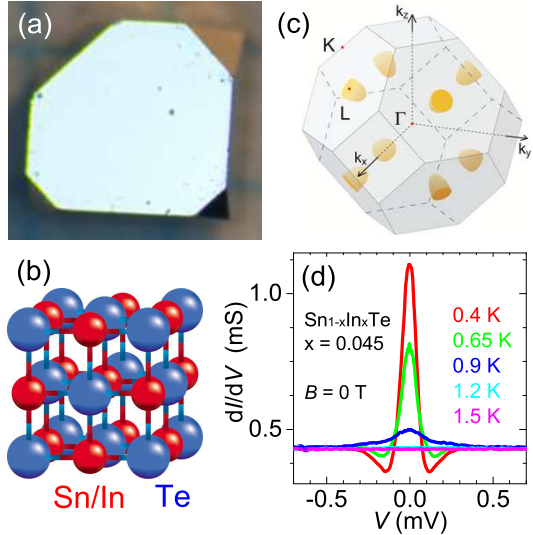}
\end{center}
\caption{Sn$_{1-x}$In$_x$Te. (a) A typical facet image of Sn$_{1-x}$In$_x$Te crystal 
grown by the vapor-transport method. (b) Cubic rocksalt crystal structure of 
Sn$_{1-x}$In$_x$Te. (c) The Fermi surfaces of p-type SnTe are centered around the 
four equivalent $L$ points, which belong to the time-reversal-invariant momenta, in 
the bulk Brillouin zone of the cubic structure with fcc Bravais lattice. 
(d) Point-contact spectra (Bias-voltage ($V$) dependence of the differential 
conductance, [$dI/dV$]($V$), at various temperatures) of Sn$_{1-x}$In$_x$Te with 
$x = 0.045$ measured in 0 T.}
\label{fig:fig4}
\end{figure}

Upon substitution of Sn by In, however, $T_c$ jumps as high as $\sim $1 K and 
gradually increases as the In content increases, suppressing the SPT, although 
the hole density remains similar \cite{Erikson}. Above $x \!\sim\!0.04$ the SPT 
is completely suppressed. Nevertheless, the topological surface state of SnTe is 
preserved on Sn$_{1-x}$In$_x$Te for $x=0.045$ providing evidence for 
the presence of an inverted band structure as in pristine SnTe~\cite{TSato13} 
and, therefore, Sn$_{1-x}$In$_x$Te with a moderate In content is a superconducting 
doped TCI. Although it has not been clarified yet whether or not Sn$_{1-x}$In$_x$Te 
with very high In content can still be a doped TCI, interestingly, progressive In 
doping enhances $T_c$ up to 4.5 K for $x=0.45$~\cite{Zhong} or $\sim $4.8 K for 
$x=0.45$ \cite{Saghir} determined by the onset of diamagnetism, and 4.7 K for 
$x \!\simeq\! 0.40$ determined by the onset of superconducting transition in the 
resistivity measurement~\cite{Balakrishnan}.

The bulk superconductivity of Sn$_{1-x}$In$_x$Te was confirmed by the specific heat 
measurement for $0.021 \!<\! x \!<\! 0.12$ \cite{Erikson}, $0.025 \!<\! x \!<\! $ 
$\sim $0.05 \cite{Novak} and $x=0.4$ \cite{Balakrishnan}, as well as muon-spin 
rotation or relaxation ($\mu$SR) measurement for $0.38 \!<\! x \!<\! 0.45$ 
\cite{Saghir}. Both measurements show the superconducting state of Sn$_{1-x}$In$_x$Te 
is likely to be fully gapped. The volume fraction of Sn$_{1-x}$In$_x$Te is 
essentially $100\%$. The absence of impurity phase is an advantage of this 
material over Cu$_x$Bi$_2$Se$_3$, if this is really a TSC.

It turns out that the superconducting state of Sn$_{1-x}$In$_x$Te has a rich phase 
diagram where the dependence of $T_c$ on indium content $x$ below $\sim $0.04 (in 
the ferroelectric rhombohedral phase) in vapor-grown (VG) single crystals \cite{Novak} 
appears to be different from that in MG single crystals \cite{Erikson}. The different 
trend of $T_c$ can be explained by the unusual role of impurity scattering within the 
framework of the BCS theory based on the Martin-Phillips theory for the enhancement 
of $T_c$ \cite{Novak} implying that the strength of the impurity scattering in crystals 
is probably different when grown in different ways. Hence, the superconducting state of 
Sn$_{1-x}$In$_x$Te in the ferroelectric rhombohedral phase is not necessarily 
unconventional and can be topologically trivial with conventional even-parity pairing.

Importantly, the ZBCP, signature of the surface ABS was observed by the soft 
point-contact spectroscopy on VG Sn$_{1-x}$In$_x$Te single crystals for $x=0.045$ as 
plotted in Fig. 4(d)\cite{SasakiPRL12}, indicating the realization of a topological 
superconducting state in the cubic phase. To identify the nature of the superconducting 
state in Sn$_{1-x}$In$_x$Te, the conduction and valence bands in the vicinity of four 
$L$ points of the fcc Brillouin zone, where ellipsoidal Fermi surfaces in the normal 
state are located as drawn in Fig.~4(c), are described by the $k\cdot p$ Hamiltonian 
in Eq. (\ref{eq:hfinal}) \cite{hsiehNP2012}. The effective Hamiltonian of 
Sn$_{1-x}$In$_x$Te has essentially the same form and the same symmetry classification 
of the possible gap functions as that of Cu$_x$Bi$_2$Se$_3$ except for the presence of 
the threefold rotation symmetry around (111) axis in Sn$_{1-x}$In$_x$Te. Theoretical 
considerations lead to the conclusion that all possible superconducting states are 
topologically nontrivial \cite{SasakiPRL12}.

However, even in the cubic phase, no clear bulk unconventionality was observed in 
Sn$_{1-x}$In$_x$Te crystals with a very high In content so far \cite{Zhong,Saghir, 
Balakrishnan}. It is worth noting that the recent numerical calculations using a self
-consistent $T$-matrix approach in the case of $k$-independent pairing reveal that the 
superconducting state can be altered from $p$- (odd-parity pairing) to $s$-wave (even
-parity pairing) character depending on the magnitude of the relativistic effects in 
the normal-state Dirac Hamiltonian of three-dimensional TSCs \cite{NagaiPRB,NagaiCon14}. 
In this regard, $s$-wave superconductivity which is robust against a strong impurity 
scattering would dominate the superconducting state of high-In-content samples where 
$\mu $ is separated from the Dirac points. Hence, the odd-parity state can be realized 
only in the lowest $T_c$ samples in the cubic phase where $\mu$ is close to the Dirac 
points and the impurity scattering is the weakest, though the surface states are 
significantly broadened due to strong quasiparticle scattering caused by In doping 
\cite{TSato13}.

Recently the effect of the In concentration on the crystal structure and 
superconducting properties of another superconducting doped TCI 
(Pb$_{0.5}$Sn$_{0.5}$)$_{1-x}$In$_x$Te has been investigated \cite{ZhongPST}. The 
single crystal of (Pb$_{0.5}$Sn$_{0.5}$)$_{1-x}$In$_x$Te grown by a modified 
floating-zone method retains the rocksalt structure up to the solubility limit of 
In ($x \simeq 0.30$). The dependence of $T_c$ and the upper critical magnetic 
field ($H_{c2}$) on the In content $x$ has been measured and the maximum 
$T_c = 4.7$ K for $x = 0.30$ with $\mu_{0}H_{c2}$ ($T =0$) $\sim $5 T was found. 
Further studies on the material to find a signature of unconventional 
superconductivity are necessary to conclude that 
(Pb$_{0.5}$Sn$_{0.5}$)$_{1-x}$In$_x$Te is a TSC. It is also important to search 
for other candidates for TSCs based on superconducting doped TCI.

\begin{figure}[t!]
\begin{center}
\includegraphics[width=50mm]{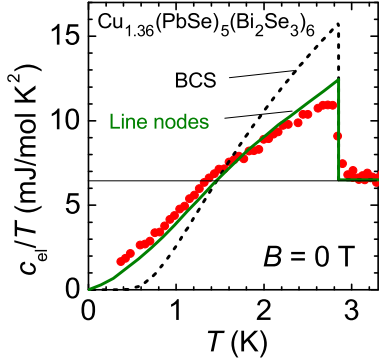}
\end{center}
\caption{Specific heat of Cu$_x$(PbSe)$_5$(Bi$_2$Se$_3$)$_6$. Superconducting transition 
in $c_{el}/T$ in 0 T obtained after subtracting the phonon contribution determined in 2 T, 
where $c_{el}$ is the electronic specific heat. The dashed line is the weak-coupling BCS 
behavior (coupling constant $\alpha = 1.76$) for $T_c$ of 2.85 K. The green solid line is 
the theoretical curve for $d$-wave pairing on a simple cylindrical Fermi surface with line 
nodes along the axial direction \cite{Maki}. Horizontal solid line corresponds to the 
normal-state electronic specific-heat coefficient $\gamma_\textsc{N}$.}
\label{fig:fig5}
\end{figure}

\subsection{Electron-doped natural heterostructure material: Cu$_x$(PbSe)$_5$(Bi$_2$Se$_3$)$_6$}

Recently, a new SC based on a topological insulator heterostructure material, 
Cu$_x$(PbSe)$_5$(Bi$_2$Se$_3$)$_6$ (abbreviated CPSBS), was discovered \cite{Sasaki14}. 
The pristine material (PbSe)$_5$(Bi$_2$Se$_3$)$_6$ (PSBS) is a natural heterostructure 
of a TI (Bi$_2$Se$_3$) and an ordinary insulator (PbSe). It was found that the PbSe unit 
works as a block layer and the topological boundary states are encapsulated in each 
Bi$_2$Se$_3$ unit, making the system to possess quasi-two-dimensional states of topological 
origin throughout the bulk \cite{Nakayama}. CPSBS is synthesized by intercalating Cu into 
PSBS with post annealing that is essential to activate the superconductivity similar to 
the activation process of Cu$_x$Bi$_2$Se$_3$. It is worth noting that the specific-heat 
behavior of CPSBS suggests for the first time in a TI-based SC that unconventional 
superconductivity occurs in the bulk with gap nodes as indicated in Fig. 5. The existence 
of gap nodes in a strongly spin-orbit coupled SC gives rise to spin-split ABSs that are 
hallmark of topological superconductivity. Hence this new SC emerges as an intriguing 
candidate for TSCs.

\section{Outlook}

\subsection{Pairing symmetry of Cu$_x$Bi$_2$Se$_3$}
\label{sec:pair}

\begin{figure}[t!]
\includegraphics[width=85mm]{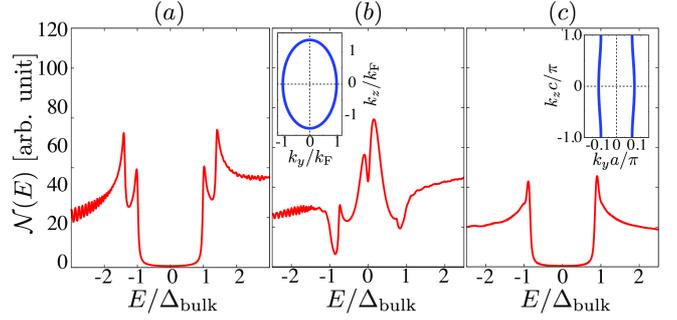}
\caption{(a) Surface density of states for the bulk $s$-wave pairing ($A_{1g}$) at $(\tilde{m}_1,\tilde{m}_2)=(-0.17,-0.20)$.  Surface density of states for the $A_{1u}$ state with a spheroidal Fermi surface (b) and a cylindrical shape (c). The insets in (b) and (c) show the Fermi surface, where $a$ and $c$ are the lattice constants.}
\label{fig:fig6}
\end{figure}

Here, we revisit the pairing symmetry and topological superconductivity of Cu$_x$Bi$_2$Se$_3$. As shown in Sec.~3.1, surface sensitive experiments observed pronounced ZBCPs on the (111) surface~\cite{sasaki11,kirzhner12,Chen12}, which are strong signatures of bulk topological odd parity superconductivity. On the other hand, conflicting experimental results of conductance and tunneling spectroscopy were recently reported~\cite{peng13,Levy13,NOTE}, which led to contrary statement that this material has a conventional $s$-wave pairing symmetry.

First of all, we would like to emphasize that, even if the bulk of Cu$_x$Bi$_2$Se$_3$ is a conventional $s$-wave pairing, i.e., the $A_{1g}$ state ($\Delta _{1a}$) in Table~\ref{table1}, the surface structure becomes unconventional~\cite{Mizushima}. As clarified in Eq.~(\ref{eq:dirac}), the Dirac fermion is fully polarized in the orbital space, which breaks the inversion symmetry. Since Cooper pairs are equally populated to both orbitals in the $A_{1g}$ state, the orbital polarization of the Dirac fermion strongly suppresses the condensation energy at the surface. As demonstrated in Ref.~\cite{Mizushima}, the mixing of the subdominant component $\Delta _3 \sigma _z$, which has the odd parity pairing, is indispensable for the maximum gain of the condensation energy. When the surface Dirac fermions that are remnants of the parent TIs are well defined, therefore, the intertwining with bulk superconductivity gives rise to a large energy gap in the surface Dirac cone, $\Delta _{\rm surf}$. This is distinguishable from the gap in the bulk conduction band, $\Delta _{\rm bulk}$. This results in the double-coherence structure of the surface density of states, as shown in Fig.~\ref{fig:fig2}(a). Hence, the bulk conventional even-parity scenario is inconsistent with a simple U-shaped form in the surface density of states reported in Ref.~\cite{Levy13} and the STS result in Sec.~\ref{sec:CBS} is a puzzle.

A possible scenario for Cu$_x$Bi$_2$Se$_3$ is bulk odd parity superconductivity with a Fermi surface evolution. For bulk odd parity pairing, no surface parity mixing is induced by Dirac fermions. The surface density of states has a pronounced zero energy peak which is responsible for a ZBCP~\cite{yamakage12,hao11,hsieh12,Takami}, as shown in Fig.~\ref{fig:fig6}(b), when the Fermi surface encloses the $\Gamma$ point. As discussed in Sec.~\ref{sec:odd}, the topological superconductivity and the existence of zero energy density of states on the surface are sensitive to the shape of the Fermi surface. For a cylindrical Fermi surface which does not enclose the $\Gamma$ point, the surface state is no longer topologically protected and thus the resultant surface density of states on the (111) surface becomes a simple U-shaped form (Fig.~\ref{fig:fig6}(c)). This feature is commonly applicable to the $A_{1u}$ and $E_{u}$ states. 

We would like to notice that the $E_u$ state is the two-dimensional representation and the arbitrary linear combination, $\sigma _y s_x \cos\phi + \sigma _y s_y\sin\phi$, rotates the nodal direction, where ${\phi}$ is the azimuthal angle and the nodal direction is denoted by the unit vector $(\sin\phi,\cos\phi)$. Since the point nodes in the $E_u$ state is protected by the mirror reflection symmetry (\ref{eq:mirror}) with respect to the $y$-$z$ plane, the bulk excitation in the $E_u$ state becomes gapful for ${\phi} \!\neq\! 0$, $\pm \pi /3$, and $\pm 2\pi/3$. Fu~\cite{fu14} recently suggested that in the presence of the extra ``warping'' term in Eq.~(\ref{eq:hfinal}), the $E_u$ scenario is able to explain all experimental results including both the bulk and surface measurements. The fully gapped $E_u$ state may be energetically competitive to the $A_{1u}$ state. 

We also mention that in the presence of a phonon-mediated short-range interaction, bulk odd parity pairing is energetically competitive to even parity pairing ($A_{1g}$). Brydon {\it et al.} demonstrated that in addition to the phonon-mediated interaction, a repulsive electron-electron interaction based on the Coulomb pseudopotential is critical to stabilizing the spin-triplet odd-parity states~\cite{brydon2014}. The bulk odd-parity pairing state, the $A_{1u}$ or fully gapped $E_u$ state, may be stabilized by phonon mechanism in the presence of a weak electron-electron correlation.

Apart from ${\bm k}$-independent pairing, momentum-dependent pairing was examined in Ref.~\cite{haoPRB2014,chenJPCM2013}. Recently, first-principle linear-response calculations predict that a $p$-wave-like state can be favored by a conventional phonon-mediated mechanism~\cite{wan14}. This is attributed to a singular behavior of the electron-phonon interaction at long wavelengths.

\subsection{Smoking-gun for topological superconductivity}
\label{sec:smokinggun}

As discussed in Sec.~\ref{sec:pair}, a promising candidate for Cu$_x$Bi$_2$Se$_3$ is an odd parity pairing, the $A_{1u}$ or $E_u$ state. Since both odd parity pairings have a twisted Majorana cone, they can equally explain a pronounced ZBCP in tunneling spectroscopy. Thus, it is hard to identify the bulk pairing symmetry via point contact measurements. Here, we summarize smoking-gun experiments for identifying bulk pairing symmetry and for extracting properties inherent to  topological superconductivity.

{\it Spin susceptibility.---}
First, the Knight shift in nuclear magnetic resonance experiments can provide a smoking-gun evidence for the spin state of the bulk pairing symmetry~\cite{hashimoto2013,hashimoto14,Zocher}. Hashimoto {\it et al.}~\cite{hashimoto2013,hashimoto14} introduced the ${\bm d}$-vector that describes the spin-triplet component of $\Delta$ in the band representation. It turns out that the Knight shift in the $A_{1u}$ state is distinguishable from that of the $E_u$ state, when an applied field is rotated within the $ab$ plane. The former is characterized by an hedge-hog-like ${\bm d}$-vector profile in the ${\bm k}$-space, similarly with that of the BW state in $^3$He. This is responsible for an isotropic magnetic response. In contrast, the $E_u$ state results in the uniaxial anisotropy of the Knight shift in the $ab$ plane. Furthermore, Nagai predicted that nodal structure in the $E_u$ state is detectable through angle-resolved heat capacity and thermal conductivity measurements~\cite{nagaiJPSJ14}. In the $E_u$ state, the zero-energy density of states around a vortex core shows twofold rotational symmetry and splits along the nodal direction with increasing energy. 

{\it Thermal Hall conductivity.---}
One of fingerprint experiments for bulk topological superconductivity is the quantized thermal Hall conductivity. The topology of three-dimensional fully gapped SCs with time-reversal symmetry is characterized by the $\mathbb{Z}$ topological number, $w_{\rm 3d}$, as defined in Eq.~(\ref{eq:windingQ}). It has been shown that the thermal Hall conductivity $\kappa _{xy}$ is quantized in $\mathbb{Z}$ time-reversal invariant TSCs, $\kappa _{xy} = \frac{\pi^2k^2_{\rm B}T}{12 h}w_{\rm 3d}$, when a small gap is induced in the Majorana cone~\cite{wangPRB11,ryuPRB12,nomuraPRL12,shiozakiPRL13,shiozakiPRB14}. Shimizu and Nomura~\cite{shimizu} evaluated the thermal Hall conductivity in the $A_{1u}$ state of Cu$_x$Bi$_2$Se$_3$. They demonstrated that $\kappa _{xy}$ depends on a time-reversal breaking perturbation which induces a small gap in the Majorana cone and $w_{\rm 3d}$ is directly detectable through the thermal Hall conductivity when the small gap is induced by the complex $s$-wave pair potential. 

{\it Topological ``crystalline'' superconductivity.---}
Topological superconductivity peculiar to nodal SCs, i.e., the $E_u$ state, is unveiled by introducing the combined ${\bm Z}_2$ symmetry which is composed of the mirror reflection symmetry and time-reversal symmetry (\ref{eq:trs}). The superconducting state retains the mirror symmetry if the gap function $\Delta({\bm k})$ is even or odd under the mirror reflection, $M\Delta({\bm k})M^{\rm T} = \pm \Delta (-k_x,k_y,k_z)$. Then, the BdG Hamiltonian $\mathcal{H}_0({\bm k})$ preserves the mirror reflection symmetry 
\beq
\mathcal{M}^{\pm}\mathcal{H}({\bm k})\mathcal{M}^{{\pm}\dag} = \mathcal{H}(-k_x,k_y,k_z),
\label{eq:mirror}
\eeq
when an external field is absent. The mirror reflection operator $\mathcal{M}^{{\pm}}$ is defined as $\mathcal{M}^{\pm} \!=\! {\rm diag}(M,\pm M^{\ast})$.

Once the Zeeman fields, ${\bm H}$, are applied, the mirror symmetry with respect to the plane parallel to the Zeeman filed is lost and the time-reversal symmetry is explicitly broken, but a combination of them can be still preserved if ${\bm H}\cdot\hat{\bm x} \!=\! 0$. This is because the combination of the mirror reflection and the time-reversal rotates the magnetic field ${\bm H}\!\rightarrow\!(-H_x,H_y,H_z)$. Consequently, the Hamiltonian $\mathcal{H}({\bm k})$ with $H_x = 0$ holds the following ${\bm Z}_2$ symmetry,
\beq
\mathcal{T}\mathcal{M}^{\pm}\mathcal{H}({\bm k})
\mathcal{M}^{\pm \dag}\mathcal{T}^{-1} = \mathcal{H}(k_x,-k_y,-k_z).
\label{eq:z2}
\eeq
Combining the ${\bm Z}_2$ symmetry with the particle-hole symmetry, 
$\mathcal{C}\mathcal{H}_{\rm eff}({\bm k})\mathcal{C}^{-1} = -
\mathcal{H}^{\ast}_{\rm eff}(-{\bm k})$, 
we define the chiral symmetry operator, 
$\Gamma _1 \!=\! \mathcal{C}\mathcal{T}\mathcal{M}^{\pm}$. Then, it turns out that $\Gamma _1$ is anti-commutable with the effective Hamiltonian
$\left\{ \Gamma _1, \mathcal{H} (0,k_y,k_z)\right\} \!=\! 0$.
Therefore, similarly to $w_{\rm 3d}$ in Eq.~(\ref{eq:windingQ}), the one-dimensional winding number is defined as
$w_{1d}(k_y) \!=\! - \frac{1}{4\pi i}\int d k_z[ 
\Gamma \mathcal{H}^{-1}({\bm k})\partial _{k_z}\mathcal{H}({\bm k})
]_{k_x=0}$~\cite{mizushimaJPCM2014,satoPRB2011,mizushimaPRL2012}.
From the generalized index theorem in Ref.~\cite{satoPRB2011}, the non-zero value of $w_{1d}$ is equal to the number of zero energy states that are bound to the surface.

Let us now consider $\Delta \!=\! \Delta _{4x}\sigma _y s_x$ in the $E_u$ state, where the point nodes lie in the mirror plane ($\phi = 0$). In this situation, one finds 
$w_{\rm 1d}(k_y) \!=\! 2 $ for $|k_y|<k_{\rm F}$, and otherwise $w_{\rm 1d}(k_y) \!=\! 0$.
There appear doubly-degenerate zero energy states along the chiral symmetric plane $k_x \!=\! 0$ in the $E_u$ state. Hence, the mirror-symmetry-protected topological invariant ensures the existence of surface Fermi arc that connects two point nodes. Such a surface Fermi arc can also be realized in the planar state of $^3$He~\cite{mizushimaJPCM2014} and the $E_{1u}$ scenario of the heavy fermion SC UPt$_3$~\cite{tsutsumiJPSJ2013} (see Fig. 7).

\begin{figure}[t!]
\begin{center}
\includegraphics[width=75mm]{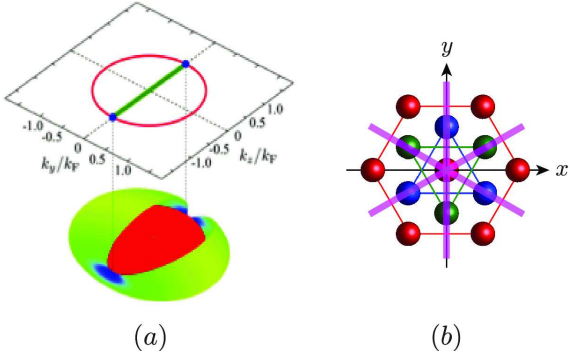}
\end{center}
\caption{(a) A stereographic view of the energy gap of the $E_u$ state. The Fermi surface and point nodes are projected onto the surface in the $k_xk_y$ plane, where the topologically protected surface Fermi arc connects to two point nodes. (b) Crystal structure of Cu$_x$Bi$_2$Se$_3$ viewed from the $c$-axis and three mirror planes.}
\label{fig:planar}
\end{figure}

Now following Refs.~\cite{mizushimaJPCM2014, mizushimaPRL2012, mizushimaNJP2013, shiozakiPRB2014}, one can show that multiple Majorana zero modes with chiral symmetry ensures the Ising character of the topologically protected zero energy states. The topologically protected surface Fermi arc does not contribute to the local density operator,
$\rho^{({\rm surf})} ({\bm r})  \!=\! 0$.
This indicates that the MFs protected by the chiral symmetry can not be coupled to the local density fluctuations and thus are very robust against non-magnetic impurities. Similarly, the local spin operator, ${\bm S}$, is constructed from the surface MF as 
${\bm S}^{({\rm surf})} \!=\! (S_x,0,0)$. This implies that only the surface MF has anisotropic magnetic response. The surface MF and topological Fermi arc are insensitive to a magnetic field along the chiral symmetric plane, but fragile against a field perpendicular to the plane. Since the surface Fermi arc is responsible for a ZBCP, the anisotropic magnetic response of MFs might be observable in point contact experiments when an applied field is rotated within the $ab$ plane. The anisotropic magnetic response can be a manifestation of TSCs with the combined ${\bm Z}_2$ symmetry, i.e., the $E_u$ state with point nodes.

We would like to notice that arbitrary linear combination, $\sigma _y s_x \cos\phi + \sigma _y s_y \sin \phi$, cannot be invariant under the mirror reflection symmetry $M \!=\! is_x$, when $\phi\!\neq\! 0$, $\pm \pi /3$, and $\pm 2\pi/3$. As mentioned in Sec.~\ref{sec:pair} and \cite{fu14}, the $E_{u}$ state without mirror symmetry becomes fully gapped. In this situation, $w_{\rm 1d}$ introduced above is irrelevant and the argument on Majorana Ising spins is not applicable to the $E_u$ state where the nodal direction is tilted from the mirror invariant plane. 

{\it Josephson coupling.---}
The existence of the Majorana cone twisted by the surface Dirac fermions is peculiar to superconducting doped topological materials. It has been predicted~\cite{Fu10PRL105, yamakage13} that bulk odd parity pairing and surface helical MF are detectable through Josephson currents. In a junction between a {\it conventional} $s$-wave and the $A_{1u}$ state of the superconducting doped topological materials, the leading order of the Josephson current $J(\varphi)$ becomes the second-order form, $J(\varphi) \!\sim\!\sin 2\varphi$, where $\varphi$ is the relative phase difference~\cite{yamakage13}. This is because a $s$-wave pairing is even under the mirror reflection, while the $A_{1u}$ state is odd as shown in Table~\ref{table1}. In addition, Yamakage {\it et al.}~\cite{yamakage13} pointed out that in the junction between superconducting TIs, the Josephson current is highly sensitive to the difference in the spin-helicity of the surface MFs at the interface, where the positive (negative) helicity characterizes the existence of Majorana ``cone'' (``caldera''). The Josephson current between different helicity is strongly suppressed at low temperatures, which is contrast to that in the junction with same helicity.

\section{Summary}

The search for Majorana fermion is currently one of the hottest issues in 
the physics community and various types of promising platforms have been 
theoretically proposed, as well as experimentally investigated during the 
last several years. In this review we focused on studies of superconducting 
doped topological materials that preserve the time-reversal-invariance. We 
began with a theoretical description of odd-parity pairing systems from a 
basic concept of symmetry invariances which are important to understand 
topological superconductors. The normal state of time-reversal-invariant 
topological superconductors is characterized by the presence of Dirac fermions. 
We discussed the role of the Dirac fermions in the topological superconductors. 
To clarify the conditions for realizing topological superconductivity in doped 
topological materials, topological invariants and possible paring symmetry in 
the system are denoted. We continued with a summary of studies on the properties 
of both normal and superconducting states in real existing materials: 
Cu$_x$Bi$_2$Se$_3$, Sn$_{1-x}$In$_x$Te, (Pb$_{0.5}$Sn$_{0.5}$)$_{1-x}$In$_x$Te, 
and Cu$_x$(PbSe)$_5$(Bi$_2$Se$_3$)$_6$. We found that the recent theoretical 
developments allow us to consistently interpret controversial results aroused 
from different experimental methods. In this review, we shed light on the 
effects of disorder in the system. In particular, we discussed that the observed 
moderate suppression of $T_c$ of Cu$_x$Bi$_2$Se$_3$ with increasing the 
concentration of Cu, which behave as non-magnetic impurities in the 
superconductor, can be explained by anomalous robustness of superconductivity 
in the presence of nonmagnetic impurity scattering, which is peculiar to 
superconducting topological materials. Finally, in order to experimentally 
identify the existence of Majorana fermions in candidate platforms of 
topological superconductors, we proposed future experiments to observe the 
anisotropic magnetic response of the candidate materials, as well as Josephson 
coupling effects.

\appendix

We thank A. A. Taskin, D. Derks, Y. Nagai, A. Yamakage, M. Sato, Y. Tanaka, and Y. Ando 
for fruitful discussions. This work is supported by JPSJ (Grants Nos.~25287085 
and ~25800199), the ``Topological Quantum Phenomena'' Grant-in Aid (No.~22103005) 
for Scientific Research on Innovative Areas from MEXT of Japan, Inamori Foundation, 
and the Murata Science Foundation.


\bibliographystyle{elsarticle-num} 
\bibliography{topo}



\end{document}